\documentclass[showpacs,twocolumn]{revtex4}

\usepackage{amsmath}
\usepackage{graphicx}

\begin{document}

\title{Quasispherical gravitational collapse  in 5D Einstein-Gauss-Bonnet gravity}
\author{Sushant G. Ghosh\footnote{sgghosh@gmail.com}
and S. Jhingan\footnote{sanjay.jhingan@gmail.com}}
\affiliation{Center for Theoretical Physics, Jamia Millia Islamia,
New Delhi 110025, India}
\begin{abstract}
We obtain a general  five-dimensional quasispherical collapsing
solutions of irrotational dust in Einstein gravity with the
Gauss-Bonnet combination of quadratic curvature terms. These
solutions are generalization, to Einstein-Gauss-Bonnet gravity,  of
the five-dimensional quasispherical Szkeres like collapsing
solutions in general relativity. It is found that the collapse
proceed in the same way as in the analogous spherical collapse,
i.e., there exists a regular initial data such that the collapse
proceed to form naked singularities violating cosmic censorship
conjecture. The effect of Gauss-Bonnet quadratic curvature terms on
the formation and locations of the apparent horizon is deduced.
\end{abstract}

\maketitle

\tolerance=5000
\newpage

\section{Introduction}
One of the remarkable special classes of the exact solution  of the
nonvacuum Einstein equations $G_{ab} = \kappa T_{a b}$ available to
date was discovered by Szekeres \cite{psz}.  They are obtained by
solving Einstein equations with irrotational dust $T_{ab} = \epsilon
u_{a} u_{b}$ as the source,  for the line element
\begin{equation}\label{qs4d}
ds^2 = -dt^2 + X^2 dr^2 + Y^2  (d x^2 + dy^2) ,
\end{equation}
relative to comoving coordinates. Here $u_a$ is a velocity (i.e.
unit timelike) vector field and $\epsilon$ is the energy density of
the system.   One can say that Szekeres solutions are obtained when
the spherical symmetry orbits in the  Lemaitre-Tolman-Bondi (LTB)
model \cite{ak1} are made nonconcentric to destroy the symmetry, but
the energy-momentum tensor is still that of dust.  Therefore, the
Szekeres spacetime is often called as quasispherical, for
definiteness we shall name it as quasispherical Szekeres (QSZ)
solutions. Being an exact model of the spacetime geometry, the QSZ
solutions have primarily been adapted with regards to studies of
nonspherical collapse of inhomogeneous dust cloud \cite{psz,psj,
jk,djj,nud,ud2,smcg} in four dimensions (4D), and in higher
dimensions \cite{ud,ud1}, in cosmology \cite{wb,gw,wb1,ck,pk,bkh}
and also in observational cosmology \cite{kb}. They are very
important because they admit no  Killing vectors \cite{wb} and hence
they are lacking symmetry.

In any attempt to perturbatively quantize gravity as a field theory,
higher-derivative interactions must be included in the action. Such
terms also arise in the effective low-energy action of string
theories. Among the higher curvature gravities, the most extensively
studied theory is the so-called Einstein-Gauss-Bonnet (EGB) gravity
\cite{bd,egbbh,jtw,rcm1,rgc,tk,hm,tmgb,gd,dg,maeda,jg}. The EGB
gravity is a special case of Lovelocks' theory of gravitation, whose
Lagrangian contains just the first three terms. Gauss-Bonnet gravity
provides one of the most promising frameworks to study curvature
corrections to the Einstein action in supersymmetric string
theories, while avoiding ghosts and keeping second order field
equations.

In Einstein-Gauss-Bonnet gravity less number of exact solutions have
been known so far. Static and spherically symmetric black hole
solutions with or without a cosmological constant, as well as a
topological black hole in an anti-de-Sitter  spacetime were obtained
\cite{bd,egbbh,jtw,rcm1,rgc}.  The effects of Gauss-Bonnet terms on
the Vaidya solutions have been investigated in
\cite{tk,hm,tmgb,gd,dg}, and on the LTB solutions in
\cite{maeda,jg}.  Here, we consider the five-dimensional (5D) action
with the Gauss-Bonnet terms for gravity and give an {\em exact
model} of the quasispherical gravitational collapse. Using our new
solution, we investigate the nature of singularities of such a
spacetime in terms of its being hidden within a black hole, or
whether it would be visible to outside observers and compare it with
analogous relativistic case.  Thus, the aim of this paper is to
extend the previous studies on the quasispherical gravitational
collapse of inhomogeneous dust, including the second order
perturbative effects of quantum gravity solutions: a Gauss-Bonnet
generalization of the QSZ solutions in 5D spacetime, namely,
5D-QSZ-EGB.

There are several issues that motivate our analysis: how does the
Gauss-Bonnet term affect the final fate of collapse?  What is the
horizon structure in the presence of the second order perturbative
effects of quantum gravity? Whether such solutions lead to naked
singularities? Do they get covered due to departure from spherically
symmetry? Does the nature of the singularity changes in a more
fundamental theory preserving cosmic censorship \cite{rp}?

\section{Szekeres Solutions in 5D  Einstein Gauss-Bonnet Gravity}
In this section, we derive the relevant equations for Szekeres model
for irrotational dust to the 5D  Gauss-Bonnet extended Einstein
equation -- 5D-QSZ-EGB solutions.   We begin with the following 5D
action:
\begin{equation}\label{action}
S = \int d^5 x \sqrt{-g} \left[\frac{1}{2\kappa_5^2}(R + \alpha
L_{GB})\right] + S_{matter},
\end{equation}
where $R$ is 5D the Ricci scalar, and $\kappa_5 \equiv \sqrt{8\pi
G_5} $ is 5D gravitational constant which is set to unity. The
Gauss-Bonnet Lagrangian is of the form
\begin{equation}\label{EGB}
L_{GB} = R^2 - 4 R_{ab}R^{ab}+R_{abcd} R^{abcd},
\end{equation}
where $\alpha$ is the coupling constant of the Gauss-Bonnet terms.
This type of action is derived in the low-energy limit of heterotic
superstring theory~\cite{Gross}. In that case, $\alpha$ is regarded
as the inverse string tension and positive definite, and we consider
only the case with $\alpha \ge 0$ in this paper.

The action (\ref{action}) leads to following set of field equations:
\begin{equation}\label{FE}
\mathcal{G}_{ab} \equiv  G_{ab} + \alpha H_{ab} = T_{a b},
\end{equation}
where
\begin{eqnarray}\label{equations}
  G_{ab} &=& R_{ab} -\frac{1}{2} g_{ab} R,
\end{eqnarray}
is the Einstein tensor and
\begin{eqnarray}
  H_{ab} &=& 2[RR_{ab}-2R_{a\alpha}R^{\alpha}_b -
  2 R^{\alpha \beta}R_{a\alpha b\beta} + R_a^{\alpha\beta\gamma}
  R_{b\alpha\beta\gamma}]  \nonumber \\& & -\frac{1}{2}g_{ab}L_{GB},
\end{eqnarray}
is the Lanczos tensor.

To find QSZ spacetime, Szekers \cite{psz} imposed no {\em a priori}
symmetry assumption, but instead postulated the metric in a special
form  which, for our 5D case, can be written as:
\begin{equation}\label{metric}
ds^2 = -dt^2 + X^2 dr^2 + Y^2 (dx^2+dy^2+dz^2) .
\end{equation}
The geodesic fluid flow vector is $\partial/ \partial t$ and the
coordinates ($x,\; y,\;z$) are comoving spatial coordinate constant
along each world-lines so that $u^a = \delta^a_t$.  $X$ and $Y$ are
functions of $(t,\;x,\; y,\; z)$ to be determined from the Einstein
equations.  The energy-momentum tensor for dust is
\begin{equation}
T_{ab} =  \epsilon(t,x,\; y,\;z) \delta_{a}^t \delta_{b}^t .
\label{eq:emt}
\end{equation}
It is seen that there exist solutions according as $Y' = 0 $ or as
$Y' \neq 0$ (throughout this paper $. \equiv \partial / \partial t$
and  $' \equiv \partial / \partial r$).  In 4D, the family of the
solution corresponding to $Y' = 0$  is a coincidental generalization
of the Friedman and Kantowski-Sachs models. Bolejko {\it et al.}
\cite{bkh} pointed out, in 4D, that the case $Y' = 0 $ found no
useful application in astrophysical cosmology. Hence, we shall
confine our discussion to the case $Y' \neq 0$. After the Einstein
equations are solved, it is required that there must exists two
functions $R(t,r)$ and $\nu(r,x,y,z)$ such  that
\begin{equation}\label{el}
Y = \frac{R(t,r)}{P(r,x,y,z)},
\end{equation}
\begin{equation}\label{ew}
X = \frac{PY'}{W(r)}.
\end{equation}
Here, $W= W(r)$ is an arbitrary function of $r$ with restriction
that $W(r)>0$.  The solution for $P$ reads
\begin{eqnarray}\label{enu}
P & = & A(r)(x^2+y^2+z^2) \nonumber \\& &  + B_x(r)^2 x+ B_y(r)^2 y
+ B_z(r)^2
 z + C(r),
\end{eqnarray}
with the free  functions $A(r)$, $B_x(r)$,  $B_y(r)$, $B_z(r)$ and
$C(r)$ satisfying an algebraic equation
\begin{equation}\label{abc}
 A C - B_x^2 + B_y^2  + B_z^2  = \frac{ \epsilon}{4} , \hspace{.2in}
 \epsilon = 0,\; \pm 1.
\end{equation}
The factor $\epsilon$ determines whether the 2-surface is spherical
($\epsilon$ = +1), pseudospherical [hyperbolic] ($\epsilon$ = -1) or
planar ($\epsilon$ = 0) \cite{ck}. Upon introducing transformations
\begin{eqnarray}
  x &=& \sin \psi \sin \phi \cot \frac{\theta  }{2}, \nonumber \\
  y &=& \cos\psi \sin \phi \cot \frac{\theta  }{2}, \nonumber \\
  z &=& \cos\phi \cot \frac{\theta  }{2} ,
\end{eqnarray}
the 3-metric yields the more familiar spherical  form
\begin{equation}\label{3-sphere}
R^2(t,r) (d\theta^2 +\sin^2\theta ( d\phi^2 + \sin^2\phi d\psi^2)) .
\end{equation}
As can be seen, if $t$ = const, and $r$= const, the above metric
becomes the metric of the three-dimensional sphere. Hence, every $t$
= const, and $r$= const slice of the Szekeres spacetime is a sphere
of radius  $R$. Thus the term quasispherical adapted.  However, the
spheres under consideration are not concentric.  Thus, the QSZ
solutions considered here are a generalization of the 5D-LTB
solutions in which sphere of constant mass are made nonconcentric.
The $A(r)$, $B_x(r)$,  $B_y(r)$, $B_z(r)$ and $C(r)$ determine how
the center of a 3-sphere changes its position in a $t=$ constant
space when the radius of the sphere is changed.  When the spheres
are concentric, the metric (\ref{metric}) becomes the line element
of the 5D-LTB model \cite{bsc,gb,gab,gds1} which arises when we set
$A=C=1/2, B_x = B_y =B_z =0 $, with no loss of generality.

The acceleration equation is given by the ${\cal G}^r_r=0$ component
of the field equations
\begin{equation}
\frac{\ddot{R}}{R}\left[1 - 4 \alpha \frac{(W^2-1)}{R^2} + 4 \alpha
\frac{\dot{R}^2}{R^2}\right] - \frac{(W^2-1)}{R^2}+
\frac{\dot{R}^2}{R^2} =0. \label{dyn01}
\end{equation}
Thus, it is necessary that $R$ satisfies the \emph{Friedmann}-like
equation
\begin{equation}
\dot{R}^2 \left[1 - 4 \alpha\frac{W^2-1}{R^2} \right] = (W^2 - 1) +
\frac{F}{R^2} - 2 \alpha \frac{\dot{R}^4}{R^2}.\label{eq:fe}
\end{equation}
 Here, $F = F(r)$ is an arbitrary function of $r$
and is referred to as mass function. In a Newtonian limit $F(r)$ is
equal to the mass inside the shell of radial coordinate $r$,
assuming that the mass function has no angular dependence. The
function $F$ must be positive, because $F < 0$ implies the existence
of negative mass. Equation~(\ref{eq:fe}) is the master equation of
the system which governs the dynamical properties of the system,
which
 is the same as one obtains in a spherically symmetric collapse of dust
in EGB \cite{jg}. Hence, we can refer to the the 5D-QSZ-EGB
discussed here as quasispherical collapsing space-times.

It is now straight forward to calculate the energy density
$\epsilon$, which we obtain
 by substituting the solutions for $X$ and $Y$. The
mass density is
\begin{equation}
\epsilon(t,x,y,z) =\frac{3}{2}\frac{P F' - 4 F P'}{P^3 X Y^3}  =
\frac{3}{2}\frac{P F' - 4 F P'}{R^3(PR'-RP')}. \label{density}
\end{equation}
This solution has in general no symmetry.  For the spherical
symmetry these solution corresponds to more conversant 5D
Tolman-Bondi models \cite{gab,gb}.

In many instance, it may be preferential to use in our work not
density $\epsilon$, but mass $M$. $ M$ is mass within sphere
($r$=constant) at given time $t$, which we compute as follows
\begin{equation}\label{mass1}
M(t,r) = \int_0^r dr \int \int \int dx\; dy\; dz\; X Y^3 \;
\epsilon.
\end{equation}
Upon using the density Eq.(\ref{density}) in (\ref{mass1}), we get
\begin{widetext}
\begin{eqnarray}
  M(t,r) &=&  \int_0^r dr\; \frac{3}{2} \int \int  \int dx\; dy\; dz\;  \frac{P F' - 4 F P'}{W P^3}\\
   &=& \int_0^r \frac{3}{2} \left[  \frac{F'}{W} \int \int \int \frac{dx\; dy\; dz\; }{P^2}
   + \frac{2F}{W} \frac{d}{dr} \int \int \int \frac{dx\; dy\; dz\; }{P^2} \right]
\end{eqnarray}
\end{widetext}
Now ${dx\; dy\; dz\; }/{P^2}$ is a metric on the unit 3-sphere, so
\begin{equation}\label{unit}
 \int \int \int \frac{dx\; dy\; dz\; }{P^2} = 2 \pi^2
\end{equation}
so that
\begin{equation}
\frac{dM}{dr} = 2 \pi^2 X Y^3 \; \epsilon
\end{equation}
and
\begin{equation}\label{mass}
  M(t,r) =  3 \pi^2 \int_0^r  \frac{F'}{W} \;  dr
\end{equation}
This is a generalization of a formula obtained for mass function in
4D QSZ \cite{psz}.  In addition, $ \epsilon > 0 $ implies $dM(r)/dr
> 0$ or ${F'(r)}{W(r)} > 0$.

It is easy to see that as $\alpha \rightarrow 0$ the master solution
(\ref{eq:fe}) of the system reduces to the corresponding 5D-QSZ
solution in \cite{ud,ud1}
\begin{equation}
\dot{R}^2 = W^2 - 1 + \frac{{F}}{R^2}. \label{eq:fe01}
\end{equation}
Rotation and acceleration of the dust source are zero, the expansion
is
\begin{equation}\label{theta}
\Theta = - \frac{\dot{Y}'}{Y'} - 3\frac{ \dot{R}}{R},
\end{equation}
and the shear  is
\begin{equation}\label{shear}
\Sigma^2 = \frac{1}{9} \left[7 \frac{\dot{Y}'}{{Y}^2} - 12
\frac{\dot{Y}'\dot{R}}{R' R}+9 \frac{\dot{R}^2}{R^2}\right].
\end{equation}
The 5D-LTB limits of these scalars can be obtained by replacing $Y$
by $R$ in the above expressions.

We assume, as in QSZ models \cite{psz},  the following regularity
conditions:- 1. The metric is $C^1$. 2. There are no shell crossing
singularities ($ Y' = PR'-RP'
> 0$). 3. The metric is locally Euclidean, i.e., we must have $W(0)
= 1.$

The dynamical properties of the model discussed above is governed by
Eq.~(\ref{eq:fe}), which is the same as one gets in the
corresponding spherical symmetric collapsing solutions \cite{jg}. To
get the 5D homogeneous dust limit, we put $W^2(r)=1+ k_0 r^2, \;
F(r) = K r^3$, where $k_0 = 0, \pm 1$ and K is constant.  Then one
gets three type of solutions - parabolic, elliptic or hyperbolic
according as $W^2(r)=1$, $W^2(r)<1$  or $W^2(r)>1$.

In particular, the 5D Friedmann-Robertson-Walker  metric follows
when $R(t,r) = S(r) f(t)$ and $W(r)=1-k_0 S^2(r)$, where $k_0$ is
the curvature index of 5D Friedmann-Robertson-Walker models and
$B_x=B_y=B_z=0,\; C=4A=1$. Then the metric (\ref{metric}) obtains
usual form
\begin{equation}\label{metric1}
ds^2 = -dt^2 + \frac{f^2 (t)}{1-k_0 S^2} dS^2 + S^2 f^2(t)
d\Omega_3^2.
\end{equation}

\subsection{Solutions for the zero-energy case, $W(r)=1$}
It may be noted that in the general relativistic case ($\alpha
\rightarrow 0$), Eq.~(\ref{eq:fe}) has three types of solutions and
$W(r) > 1$, $W(r) = 1$ or $W(r) < 1$ determines the type of
evolution.  From Eq.~(\ref{eq:fe}), we obtain
\begin{eqnarray}\label{dyn0}
{\dot R}^2 = (W^2-1) -  \nonumber \\ \frac{R^2}{4 \alpha} \left(1
\mp \sqrt{1+ \frac{16 \alpha^2}{R^4}(W^2-1)^2 + \frac{8 \alpha
F(r)}{R^4}} \right).
\end{eqnarray}
There are two families of solutions that correspond to the sign in
front of the square root in Eq.~(\ref{dyn0}). We call the family
that has the minus (plus) sign the minus (plus) branch solution. In
the general relativistic limit ${ \alpha} \to 0$,  we recover the
$5D$-QSZ solution in Einstein gravity \cite{ud,ud1}.  Maeda
\cite{maeda} has analyzed LTB models near center ($r \sim 0$) in EGB
and pointed out the occurrence of major changes in the final fate of
collapse (see also, \cite{jg}). Here we present the 5D-QSZ-EGB exact
solution in close form, which facilitates us to analyze the final
fate of gravitational collapse. The condition $W(r)=1$, is the
marginally bound condition, meaning the collapsing shell is at rest
at spatial infinity $(R=\infty)$ with zero energy in the infinite
past. In the present discussion, we are concerned with gravitational
collapse, which requires $\dot{R}(t,r) < 0$, i.e., we assume that
all portion of the dust cloud are momentarily collapsing.
Eq.~(\ref{dyn0}) can be integrated to
\begin{eqnarray}\label{solution}
{t_\varsigma(r)-t}  = {\frac{\sqrt{\alpha}}{2\sqrt{2}}} \tan^{-1}
\left[\frac{3 R^2 -\sqrt{R^4+8\alpha F}}{2\sqrt{2} R [\sqrt{R^4 +
8\alpha F} -R^2]^{1/2}} \right] \nonumber \\ +\sqrt{\frac{\alpha
R^2}{\sqrt{R^4+8\alpha F}-R^2}}\;, \quad
\end{eqnarray}
where $t_\varsigma(r)(r)$ is an arbitrary function of integration.
Here we note that the solutions $R(t,r)$ are same as the 5D-LTB-EGB
models \cite{jg} and are not affected by the dependence of the
Szekeres model on the  $(x,\;y,\;z)$ coordinates.
 As it is possible to make an arbitrary
relabeling of spherical dust shells by $r \rightarrow g(r)$, without
loss of generality, we fix the labeling by requiring that, on the
hypersurface $t = 0$, $r$ coincide with the area radius
\begin{equation}
R(0,r) = r.             \label{eq:ic}
\end{equation}
This corresponds to the following choice of $t_\varsigma(r)$:
\begin{eqnarray}\label{scale}
{t_\varsigma(r)} = {\frac{\sqrt{\alpha}}{2\sqrt{2}}} \tan^{-1}
\left[\frac{3 - \sqrt{1 + 8\alpha {\tilde F}}}{2\sqrt{2}[\sqrt{1 +
8\alpha {\tilde F}}-1]^{1/2}}\right] \nonumber \\ + \sqrt{
\frac{\alpha}{\sqrt{1+8\alpha {\tilde F}}-1}}\; , \quad
\end{eqnarray}
where $\tilde F = F/r^4$. Now, we discuss the nature and occurrence
of curvature singularities in the 5D-QSZ-EGB models. It follows from
 Eq.~(\ref{density}) that the curvature singularities occurs in
5D-QSZ-EGB solutions when
\begin{equation}\label{sing}
R =0 \; \mbox{or} \; Y' = PR'-RP' =0.
\end{equation}
Szekeres \cite{psz} analyzed the singularities of the first kind and
second kinds according to the above two conditions.  As is the case
of the LTB, in 4D, $Y'=0$ corresponds to a shell crossing, however
it is qualitatively different from that which occurs in LTB models (
see \cite{bkh} for details).  In this paper we shall restrict
ourselves to the singularities of the first kind which we call the
central singularity. Let us assume at $t = t_c(r)$, we have $R(t,r)
= 0$, which is the time when the matter shell $r$ = constant hits
the physical singularity, i.e., $t_c(r)$ is defined by $R(t_c(r),r)
= 0$.

The  central singularity curve can be obtained using Eq.
(\ref{solution}) as
\begin{equation}\label{sing-curve}
t_c(r) =t_\varsigma(r) +\frac{\pi \sqrt{\alpha}}{4\sqrt{2}},
\end{equation}
which represents the proper time for the complete collapse of a
shell with coordinate $r$. Interestingly, positive $\alpha$ delays
the formation of singularity. In the limit of vanishing $\alpha$ we
recover the crunch time for relativistic 5D-QSZ-EGB. The eight
arbitrary functions $A(r)$, $B_x(r)$,  $B_y(r)$, $B_z(r)$, $C(r)$
$F(r)$, $W(r) > 0 $ and $t_{c}(r)$ completely specify the dynamics
of collapsing shells. However, only seven are independent because
relation (\ref{abc}) is between them.  With the choice of $r$, one
can fix one more function and thus the total degree of freedom is
six. The corresponding 5D-LTB-EGB models have only three free
functions: $F(r)$, $W(r)$ and $t_{c}(r)$. Hence the 5D-QSZ-EGB is
functionally more generic.

In order to study the collapse of a finite spherical body in EGB, we
have to match the 5D-QSZ-EGB solution along the timelike surface at
some $r = r_c >0 $ to the $5D$-EGB Schwarzschild exterior discovered
by Boulware and Desser \cite{bd}, and Wheeler \cite{jtw}. The
analysis is similar to the case of matching 5D-LTB-EGB to the
$5D$-EGB Schwarzschild exterior \cite{maeda} and, will not be
discussed here. Bonnor \cite{wb} was the first to proved that
Szekeres spacetime can be matched to Schwarzschild vacuum spacetime.

\section{Apparent Horizon and Trapped Surface}
The apparent horizon (AH) is the outermost marginally trapped
surface for the outgoing photons.  The AH can be either null or
spacelike, that is, it can "move" causally or acausally.  The main
advantage of working with the apparent horizon is that it is local
in time and can be located at a given spacelike hypersurface. The
event horizon instead is nonlocal. Moreover, the event horizon is a
an outer covering surface of apparent horizon and they coincide in
case of static or stationary spacetime. Trapped surface is defined
as a compact spacelike 2-surface both of whose future pointing null
geodesics families are converging. Physically, it captures the
notion of trapping by implying that if 2-surface $S_{(r, t)}$ ($t,r$
= contant) is a trapped surface then its entire future development
lies behind a horizon. To obtain criterion for existence of such
$S_{(r, t)}$, let $K^{a}$ be the tangent vector to the null
geodesics. It follows that along null geodesics, we have
\begin{equation}\label{null_curves}
K_{\mu} K^{\mu} = 0, \qquad K^{\mu}_{; \nu} K^{\nu} = 0 ,
\end{equation}
and $K^{x} = K^{y} = K^{z} = 0$. For the metric (\ref{metric}) the
above defined null congruence satisfies the following condition:
\begin{equation}\label{null_geo}
(K^{t})^2 - X^2 (K^{r})^2 = 0 ,
\end{equation}
on $S_{(r, t)}$. A choice of affine parameter
\begin{equation}\label{affine}
K^{t} = X, \qquad K^{r} = \varepsilon,
\end{equation}
with $\varepsilon = \pm 1$, clearly satisfies the condition
(\ref{null_geo}). The positive (negative) sign of invariant
$K^{\mu}_{; \mu}$ is determines the divergence (convergence) of the
null geodesic.  Also $\pm$ corresponds to the outward and inward
geodesic respectively.  Now,
\begin{equation}\label{cov_div}
K^{\mu}_{; \mu} = K^{t}_{,t} + K^{r}_{,r} + \varepsilon
\left(\frac{X'}{X} + 3\frac{Y'}{Y} \right) + X \left(\frac{{\dot
X}}{X} + 3\frac{{\dot Y}}{Y} \right).
\end{equation}
Upon taking the time derivative of the null condition $K^{\mu}
K_{\mu} = 0$, we get
\begin{equation}\label{nul_cond_1}
K^{t}_{,t} -\varepsilon X K^{r}_{,t} - {\dot X} =0.
\end{equation}
In the second condition $K^{\mu}_{; \nu} \; K^{\nu} = 0$, consider
the $\mu =1 $ component;
\begin{equation}\label{cond_2}
K^{r}_{,r} +\varepsilon X K^{r}_{,t} + \varepsilon \frac{X'}{X} +
2{\dot X} = 0.
\end{equation}
Eliminating $K^{r}_{,t}$ from the previous two equations we can
rewrite Eq.~(\ref{cov_div}) as
\begin{equation}\label{aphor}
K^{\mu}_{;\mu} = 3 \varepsilon \frac{Y'}{Y} + 3 X \frac{{\dot
Y}}{Y}.
\end{equation}
Thus the apparent horizon must satisfy
\begin{equation}
\varepsilon Y' +X {\dot Y} = 0.
\end{equation}
In absence of shell crossings ($Y' \neq 0$), and using
Eqs.~(\ref{el}) and (\ref{ew}), we have
\begin{equation}\label{app1}
\dot{R}^2(t_{AH}(r),r) = - W(r).
\end{equation}
Considering Eq.~(\ref{eq:fe}), the apparent horizon condition
(\ref{app1}) becomes
\begin{equation}\label{app-cond}
R(t_{AH}(r),r) =  \sqrt{F(r) - 2 \alpha}.
\end{equation}
One can also employ the usual condition for the existence of
 the apparent horizon
\begin{equation}\label{app}
g^{a b} Y_{,a} Y_{,b} = 0.
\end{equation}
Considering Eqs.~(\ref{el}),(\ref{ew}) and (\ref{eq:fe}), the
apparent horizon condition (\ref{app}) again gives the same result
(\ref{app-cond}). In the relativistic limit, $\alpha \rightarrow 0$,
$R_{AH} \rightarrow \sqrt{F(r)}$ \cite{gb}. Further,
Eq.~(\ref{app-cond}) has a mathematical similarity for the analogous
situation in null fluid collapse where the expression for apparent
horizon is $r_{AH} = \sqrt{m(v) - 2 \alpha}$ \cite{gd}.

\section{End state of collapse}
In this section, we analyze end state of the collapse of 5D-QSZ-EGB
dust collapse in terms of the given regular initial density and
velocity profile. We denote $\rho(r)$ as the initial density of the
dust cloud at $t=0$ for a fixed radial direction ($x=y=z=$
constant).  It is assumed that $\rho$, $P$ and $F$ to be expandable
around $r=0$ on the initial hypersurface.  Further the function $P$
is restricted by Eqs.~(\ref{enu}) and (\ref{abc}). Henceforth, in
this section, we adopt here a method similar to \cite{psj,djj} which
we modify here to accommodate the higher dimension spacetime.  We
take
\begin{equation}
\rho = \sum_{n=0}^{\infty} \rho_n r^n,
\end{equation}
\begin{equation}
P = \sum_{n=0}^{\infty} P_n r^n, \hspace{.2in}
\end{equation}
where  $P_0 \neq 0$ and the regularity condition implies that $P_1
=0$ \cite{djj}, and
\begin{equation}\label{rho}
F = \sum_{n=0}^{\infty} F_n r^{n+4} .
\end{equation}
Therefore,
\begin{eqnarray}
  P' &=& \sum_{n=2}^{\infty} n P_n r^{n-1}, \\
  F' &=& \sum_{n=0}^{\infty} (n+4) F_n r^{n+3} .
\end{eqnarray}
Substituting $P'$ and $F'$ in the $\rho$ equation one gets,
\begin{eqnarray}\label{fc}
  F_0 &=& \frac{\rho_0}{6}, \hspace{.2in}   F_1 = \frac{2 \rho_1}{15},  \nonumber \\
  F_2 &=& \frac{\rho_2}{9}, \hspace{.2in}    F_3 = \frac{2\rho_3}{21} - \frac{4 P_2 \rho_1}{105 P_0}, \nonumber \\
  F_4 &=& \frac{\rho_4}{12} - \frac{1}{P_0} \left(\frac{ P_2 \rho_2}{18} - \frac{P_3
  \rho_1}{20},
  \right), \nonumber \\
  F_5 &=& \frac{2 \rho_5}{27} - \frac{2 1}{9 P_0}, \nonumber \\
  & & \times  \left( \frac{4 P_4 \rho_1}{15} + \frac{P_3 \rho_2}{3} + \frac{2 P_2 \rho_3}{7} + \frac{2 P_2^2 \rho_1}{105}
  \right).
\end{eqnarray}
This implies that $\rho_0, \rho_1, \rho_3$ can not have angular
dependence whereas $\rho_n$ (for $n \geq 3$) have angular dependence
provided $\rho_{n-2} \neq 0$.
 The results obtained here are consistent with the
earlier work \cite{psj,djj}, but with some changes in the
coefficients due to 5D spacetime. Since earlier analysis were done
in QSZ spacetime, as a result, one may conclude that the 5D-QSZ-EGB
spacetime has same local nakedness behavior as the QSZ spacetime. To
conserve space, we shall avoid replication of all other analysis
being similar to QSZ spacetime \cite{psj,djj}. To further analyze
the horizon curve, we combine Eqs. (\ref{solution}) and
(\ref{sing-curve}) giving
\begin{eqnarray}\label{new_dynamics}
{t_c(r)-t} &=& \frac{\pi\sqrt{\alpha}}{4\sqrt{2}} +
\sqrt{\frac{\alpha R^2}{\sqrt{R^4+8\alpha F}-R^2}} \\
\nonumber &+&{\frac{\sqrt{\alpha}}{2\sqrt{2}}} \tan^{-1}
\left[\frac{3 R^2 -\sqrt{R^4+8\alpha F}}{2\sqrt{2} R [\sqrt{R^4 +
8\alpha F} -R^2]^{1/2}} \right] .
\end{eqnarray}
The apparent horizon in the interior of the dust ball lies at
$R(t_{AH}(r),r) =  \sqrt{F(r) - 2 \alpha}$. The corresponding time
$t_{\mbox{AH}}(r)$ is given by
\begin{eqnarray}\label{sin-app}
t_c(r) - t_{AH}(r) = \frac{\pi \sqrt{\alpha}}{4\sqrt{2}}
+\frac{\sqrt{\alpha}}{2\sqrt{2}} \tan^{-1}\left[
\frac{F-4\alpha}{2\sqrt{2\alpha (F-2\alpha)}} \right] \nonumber \\ +
\frac{1}{2} \sqrt{F-2\alpha} \; . \quad
\end{eqnarray}
As mentioned above, at $t = t_c(r)$, we have $R(t,r) = 0$, which is
the time when the matter shell $r$ = constant hits the physical
singularity. The singularity is at least locally naked if $t_{AH} >
t_{c}$, and if $t_{AH} > t_{c}$, it is a black hole, and in case of
the equality one has to compare the slopes. We consider the
following three cases:
\paragraph {Case  I} Homogeneous case, i.e., the density profile is homogenous and has no angular dependence, thus
\begin{equation}\label{fhm}
F = F_0 r^4
\end{equation}
Using Eq.~(\ref{fc}), it can be deduced that $t_{c} < t_{AH}$ (see
also Fig.1), and hence singularity is naked.
\paragraph {Case  II} Next, we assume that only $F_1 \neq 0$  implies that the density profile is inhomogeneous and has no angular
dependence
\begin{equation}\label{fna}
F = F_0 r^4 + F_1 r^5
\end{equation}
It is seen that $t_{c} < t_{AH}$  which leads to formation of a
naked singularity.
\paragraph  {Case III} Finally, let $F_3 \neq 0$ which means the density profile is inhomogeneous  and has  angular
dependence
\begin{equation}\label{fwa}
F = F_0 r^4 + F_3 r^7
\end{equation}
In this case also singularity is naked because $t_{c} < t_{AH}$ (see
also Fig.1).

Clearly, for a positive $\alpha$, the central shell does not get
trapped, and the untrapped region around the center increases with
increasing $\alpha$, for both homogeneous and inhomogeneous models.
The center ($r=0$) remains untrapped, since for nonzero values of
the Gauss-Bonnet parameter $\alpha > 0$, above Eq.~(\ref{app-cond})
admits no solution.  Interestingly, the theory demands $\alpha$ to
be a positive number, which forbids the apparent horizon from
reaching the center thereby making the singularity massive and
eternally visible, which is forbidden in corresponding general
relativistic scenario.

\begin{figure}[ht]
\centering
\includegraphics[width=5.0 cm,angle=-90]{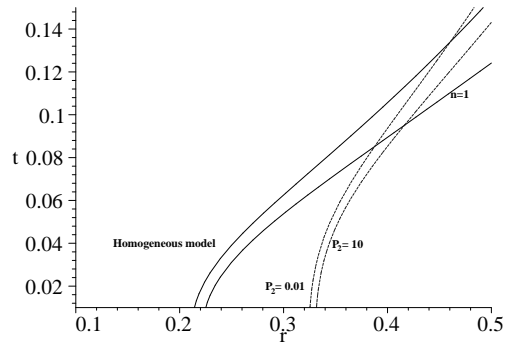}
\caption{A two-dimensional picture of the 5D-QSZ-EGB collapse
showing $ t= t_{c}(r) - t_{AH}(r)$ versus $r$ for the suitable
values of parameters. The
 continuous curves used when density has no angular dependence  and
 when density has angular dependence dotted curves are used.}
\label{figure1}
\end{figure}

\section{Effect of the Gauss-Bonnet Term }
It is seen here that the Gauss-Bonnet term modifies the time of
formation of singularity, and the time lag between singularity
formation and the apparent horizon formation, in contrast to the 5D
dust models. Indeed, the time for the occurrence of the central
shell focusing singularity for the collapse is increased in
comparison to the 5D-QSZ case. The introduction of a Gauss-Bonnet
term slows down the collapse process \cite{jg}, which can be
explained as follows. The contribution from the Gauss-Bonnet term
leads, when we think in terms of a Newtonian potential, to a
repulsive term. The Eq.~(\ref{eq:fe}), for the marginally bound
case, becomes
\begin{equation}\label{eq:fe1}
{\dot R}^2  = \frac{F}{R^2} - 2 \alpha \frac{\dot{R}^4}{{R^2}} .
\end{equation}
 In the $\alpha \rightarrow
0$ limit we recover the 5D-QSZ solution, and in this limit,
Eqs.~(\ref{eq:fe}) and (\ref{eq:fe1}) lead to the following
expression for acceleration:
\begin{equation}\label{NForce}
\ddot{R}_{5D-QSZ} = - \frac{F}{R^3} .
\end{equation}
In the 5D-QSZ-EGB spacetime the acceleration expression
(\ref{eq:fe}) can now be expressed as
\begin{equation}\label{acc3}
\ddot{R}_{5D-QSZ-EGB}= -\frac{R}{4\alpha} + \frac{R}{4\alpha}
\left[1 + 8\alpha \frac{F}{R^4}\right]^{-1/2} .
\end{equation}
For a comparison with Newton like force equation obtained in the
5D-QSZ case $\Phi(R) = -F/R^3$,  clearly a positive value of
$\alpha$ diminishes the acceleration in the collapse process. As we
move out to larger shells and for large values of $\alpha$ the
detailed dynamics should depend on higher order terms in the
expansion.

The other reason may be, for slow down of the collapse process in
5S-QSZ-EGB, there is relatively less mass energy [see
Eq.~(\ref{eq:m3})] collapsing in the 5D-QSZ-EGB spacetime as
compared to the 5D-QSZ case.  This can be seen from the mass
function $m(t,r)$, which is given by
\begin{eqnarray}
m(t,r) & = &R^2 \left(1 - g^{ab} R_{,a} R_{,b} \right) .
\label{eq:m2}
\end{eqnarray}
Using Eqs.~(\ref{el}), (\ref{ew}) and ~(\ref{eq:fe}) into
Eq.~(\ref{eq:m2}) we get
\begin{equation}
m(t,r) = F(r)  - 2 \alpha \dot{R}^4 . \label{eq:m3}
\end{equation}
The quantity $F(r)$ can be interpreted as energy due to the energy
density $\epsilon$.

Finally, for the reason mentioned above,  the presence of the
coupling constant of the Gauss-Bonnet terms $\alpha$ produces a
change in the location of these horizons. Such a change could have a
significant effect in the dynamical evolution of these horizons. For
nonzero $\alpha$ the structure of the apparent horizon is
nontrivial. In general relativity noncentral singularity is always
covered \cite{dc} (see also \cite{cjjs}). However, in the presence
of the Gauss-Bonnet term we find that even the noncentral
singularity is naked, in spite of being massive ($F(r>0) > 0$).

\section{discussion}
The Szekers models, depending on the sign of $\epsilon$, are
subdivided in to the quasispherical, quasiplane and quasihyperbolic
\cite{bkh,pk}.  The geometry of the later two is not really
understood \cite{bkh,pk}.  On the other hand the quasispherical has
 been rather well investigated \cite{wb,gw,wb1,ck,pk,bkh,kb}, and it
has found important applications in cosmology and gravitational
collapse. The QSZ metric is a dust model which has no Killing vector
\cite{wb}, but contains LTB model as a spherically symmetric special
case. It has been found that the LTB metric admits both naked
singularities and black holes depending upon the choice of initial
data. Indeed, both analytical \cite{dc,cjjs,dj,jd1,jjs} and
numerical results \cite{es} in dust indicate the critical behavior
governing the formation of black holes or naked singularities. A
similar situation also occurs in higher dimension the LTB models
\cite{bsc,gb,gab,gds1}, and these results also carry over to QSZ
spacetime \cite{psj,jk,djj}. Maeda \cite{maeda} and we \cite{jg}
have shown that in spherically symmetric inhomogeneous dust
collapse, the effect of adding a positive $\alpha$ does radically
alter the final fate and leads to formation of a massive timelike
singularity which is prohibited in general relativity.

In this work, we have obtained an exact solution in closed form in
EGB, which represents the quasispherical collapse of irrotational
dust in 5D spacetime namely 5D-QSZ-EGB. The solution is nonsymmetric
generalization of the spherically symmetric 5D-LTB-EGB solutions. It
can be reduced to  5D-QSZ or 5D-LTB in the general relativistic case
($\alpha \rightarrow 0$). Our analysis also supports the earlier
results \cite{djj}, deducing that under physically reasonable
initial conditions naked singularities do develop in the 5D-QSZ-EGB
models, which are not spherically symmetric, and admit no Killing
vectors. The second order curvature corrections changes the final
fate of gravitational collapse and the nature of singularity that
occurs in 5D general relativistic dust models.  However,  mild
departure spherical collapse can not alter the standard picture of
the structure and formation singularity, since the 5D-QSZ-EGB
solutions discussed here are qualitatively similar to the analogous
spherical solutions. It is seen here that the Gauss-Bonnet term: (i)
decelerates the collapse process (ii) alters the time of formation
of singularities and the time lag between singularity formation, and
(iii) modifies the apparent horizon formation and the location of
apparent horizons. Our analysis to examine the nature of
singularities (naked or hidden by horizon) is based on the
comparison of $t_{AH}$ (time for the formation of the apparent
horizon or trapped surface) and $t_c$ (time for the formation of a
central singularity).  In QSZ  the singularity can be directionally
naked \cite{djj}. Hence, it would be necessary to investigate in
further details the final fate of the inhomogeneous dust collapse in
 the EGB theory in order to bring out explicitly the difference in global
nakedness when we depart from spherically symmetric \cite{sj1}.

The conjecture that such a singularity from a regular initial
surface must always be hidden behind an event horizon, called cosmic
censorship conjecture (CCC) was proposed by Penrose \cite{rp}.   The
CCC forbids the existence of naked singularities.  Despite almost 30
years of effort, we are far from a general proof of CCC (for recent
reviews and references, see \cite{r1}).  But, significant progress
has been made in trying to find counter examples to CCC.  Our
analysis, as in the 5D-LTB-EGB \cite{jg}, shows that there exists a
regular initial data which leads to a naked singularity and hence in
our nonspherical case also the CCC is violated.  The usefulness of
these models is that  they do offer an opportunity to explore the
properties of singular spacetime. The investigation of a mild
departure from standard spherical symmetry models may be valuable in
attempts to put CCC in concrete mathematical form. Finally, studying
such models which lacks symmetry is important, so that one can check
which properties of gravitational collapse are preserved.  This may
also helps us to bring out some universal features in the theory of
gravitational collapse \cite{sj1}. We have shown here that there
exist regular initial data which leads to a naked singularity
violating CCC.  However, this may not be a serious threat to CCC
because of the following two reasons viz. (i) The matter considered
here is dust which is only an effective, macroscopic approximation
to a fundamental description of matter \cite{wald}. (ii) In general
relativity the energy-momentum tensor given by Eq.~(\ref{eq:emt})
satisfies the weak energy condition. However, this may not be true
in EGB because the Gauss-Bonnet term itself violates the energy
condition.

\acknowledgements It is our pleasure to thank N. Dadhich for helpful
comments and discussion.  SGG would also like to thank IUCAA, Pune
for hospitality while part of this work was done. SJ acknowledges
support under DST project No.SR/S2/HEP-002/2008.

\end{document}